# Machine Learning in Falls Prediction; A cognition-based predictor of falls for the acute neurological in-patient population

Mr. B.A. Mateen[1]

Mr. M. Bussas[2]

Dr. C. Doogan [3]

Dr. D. Waller[4]

Dr. A. Saverino[5]

Dr. F. J. Király[2,6*]

Prof. E. D. Playford[3,7]

[1] University College London, London, UK
[2] Department of Statistical Science, University College London, London, UK
[3] Therapy and Rehabilitation Services, National Hospital for Neurology & Neurosurgery, London, UK
[4] Neurorehabilitation Unit, National Hospital for Neurology and Neurosurgery, London, UK
[5] Wolfson Neuro Rehabilitation Centre, St Georges Hospital, London, UK
[6] The Alan Turing Institute, London, UK
[7] Institute of Neurology, University College London, London, UK

**Author for correspondence (*):**
Dr. Franz J. Király
Department of Statistical Science, University College London,
Gower Street London WC1E 6BT United Kingdom
Tel.: +44 - 20 - 7679 1259 Fax.: +44 - 20 - 3108 3105
E-mail: f.kiraly@ucl.ac.uk

**Declarations –**
Dr Diane Playford was supported by the National Institute for Health Research University College London Hospitals Biomedical Research Centre. The authors have no conflict of interests to declare. This study received no funding.

**Contributions –**
BAM and EDP conceived and planned the study with contribution from CD, DW and AS. BAM collected the data under the supervision of DW and EDP. Statistical analysis conducted by MB under the supervision of FJK, with contributions from BAM. Manuscript written by BAM and FJK, under the supervision of EDP.

## Abstract

**Background Information –** Falls are associated with high direct and indirect costs, and significant morbidity and mortality for patients. Pathological falls are usually a result of a compromised motor system, and/or cognition. Very little research has been conducted on predicting falls based on this premise.

**Aims –** To demonstrate that cognitive and motor tests can be used to create a robust predictive tool for falls.

**Methods –** Three tests of attention and executive function (Stroop, Trail Making, & Semantic Fluency), a measure of physical function (Walk-12), a series of questions (concerning recent falls, surgery and physical function) and demographic information were collected from a cohort of 323 patients at a tertiary neurological center. The principal outcome was a fall during the in-patient stay (n = 54). Data-driven, predictive modelling was employed to identify the statistical modelling strategies which are most accurate in predicting falls, and which yield the most parsimonious models of clinical relevance.

**Results –** The Trail test was identified as the best predictor of falls. Moreover, addition of any others variables, to the results of the Trail test did not improve the prediction (Wilcoxon signed-rank $p < .001$). The best statistical strategy for predicting falls was the random forest (Wilcoxon signed-rank $p < .001$), based solely on results of the Trail test. Tuning of the model results in the following optimized values: 68% (± 7.7) sensitivity, 90% (± 2.3) specificity, with a positive predictive value of 60%, when the relevant data is available.

**Conclusion –** Predictive modelling has identified a simple yet powerful machine learning prediction strategy based on a single clinical test, the Trail test. Predictive evaluation shows this strategy to be robust, suggesting predictive modelling and machine learning as the standard for future predictive tools.

Words - 284

# Introduction

### The Cost and Prevalence of Falls, and Falls-related Injury

Falls are a serious public health concern[1], with potentially fatal consequences[2], and significant financial implications for individuals, and their families[3]. In a single year in the USA, there were more than 10,000 fatal falls in the elderly population, and an additional 2.6 million medically treated falls-related injuries that were non-fatal, which resulted in a direct cost of close to US $20 billion[4]. In the UK, falls account for over 60% of all hospital in-patient related safety incidents[5], resulting in an annual direct cost of £15 million[6], on top of the billions already spent on treating falls-related injuries in the community that result in hospital admissions[7,8]. Some argue that if steps are not taken to address this problem, by the year 2030 the number of injuries resulting from falls will have increased by 100%[9], therefore it is vital that steps are taken to prevent this astronomical rise in cost and harm to all the relevant stakeholders.

### Clinical Relevance of Predicting Falls

The current debate in the falls literature is whether probabilistic prediction is clinically useful, and whether it is more important than targeting modifiable risk factors[10]. We argue that these two approaches are not mutually exclusive, rather, making sound predictions is actually necessary for planning and evaluating interventions of any kind, including those targeted at risk factors[11].

### State of the Art and Challenges in Predicting Falls

The STRATIFY Tool is the gold standard predictive tool for falls in geriatric patients. Although widely used in the UK, has not been improved upon in two decades[12,13], due to two notorious key issues, which we explain and address in our study:

**(I) Reproducibility** – replication studies have repeatedly failed to reproduce the reported good performance of the initial STRATIFY validation study[14]. This is due to the missing statistical evaluation in terms of expected performance on new, unseen data – which we address by **predictive modelling** and **predictive model evaluation**, including a precise quantification of expected future performance in a similar setting.

**(II) Interpretability** – it has remained unclear what the STRATIFY Tool actually measures in terms of cognitive ability or physical function. Three of the five questions record confusion, visual impairment, and frequent toileting, which, at best, are proxy measurements. We instead use **direct measures of cognitive and physical function**, such as the Trail test and the Walk-12, which are well-validated and readily interpretable.

**Paradigm Shift I: Predictive Modelling**

**Descriptive modelling**, such as in traditional linear/logistic regression analysis, aims to *fit* the data well and is powerful for identifying associations present in the data.

However, they often turn out to be *too closely* fitted to the data analyzed and do not generalize well to new, unseen data – this phenomenon is known as overfitting. We argue that this well-known phenomenon explains the supposed loss of predictiveness and accuracy in replication studies[14] better than additional, hypothetical changes in the patient collective.

On the other hand, models obtained by **predictive model selection** often contain *less variables* and *generalize more robustly*. The main difference between exploratory/descriptive and predictive modelling does not lie in the type of models applied (for example, linear or logistic regression models occur in both), but in how. The predictive modelling paradigm has led to the development of a number of non-linear modelling techniques found in the machine learning community such as kernel methods or random forests which are specifically designed to produce models that generalize well. Furthermore, the underlying theory[15] provides explicit meta-methods to quantitatively estimate accuracy on unseen data, which is used to identify the best models and most informative variables, as opposed to descriptive approaches which rely on quantifying how accurately the model describes the available data. Our proposed solution is shift into the superior modelling paradigm, and subsequent evaluation of the results in terms of their out-of-sample error/predictive error.

There are prior instances of assessing falls models via predictive evaluation[16]; however, to our knowledge, our work is the first instance where predictive model selection is employed not only to identify types of models but also the most relevant and clinically useful variables.

**Paradigm Shift II: Direct Measurements of Neuropsychological and Physical Function**

The current state-of-art in predicting falls assesses patients based on risk factors such as age, urinary urgency, or walking impairment. We argue that these risk factors are in fact **proxy measurements of cognitive and physical function**. For example, the association between age and falls, can be thought of as a result of declining executive function and attention[17], and reduced mobility[18], which both occur as we grow older. Both UTI-associated urinary urgency and the associated cognitive deficits can exacerbate the risk associated with any reductions in physical mobility[19].

Hence we propose the use of **direct measurements of cognitive and physical function** instead; a premise that the literature has alluded to increasingly often over the last few years[20]. In our study, we consider the Stroop Colour-Word test, the Trail Making test, a Semantic Fluency test as direct measurements of cognitive function, and a PROM of physical function.

# Material and Methods

### Neuropsychological Data

The test battery (described in appendix) consisted of neuropsychological tests of attention and executive function (Stroop Colour-Word tests, the Trail Making tests, and a Semantic Fluency test), a PROM (Patient Reported Outcome Measure) of motor function (Walk-12), three questions relating to past 1 month's medical history (undergone surgery; change in physical function; and, fallen over), and demographic data. Data was collected from a convenience sample (see appendix for recruitment details) of 323 patients from 3 neurosurgical, 3 neurological, and 2 neuro-rehabilitation wards, at a tertiary neuroscience center (summary statistics presented in appendix). The principal outcome in the prospective study was whether a patient fell (n = 54) or not during their in-patient stay (inclusion/exclusion criteria, cohort demographics, and summary statistics can be found in appendix). A fall was defined as a suspected, reported or witnessed incident, which consisted of unintentional contact with the ground (or intermediary object, which halted their progression to the floor, e.g. a wall), by any part of the body, except the feet. The additional distinction of recurrent falling has been disregarded in this study as a single fall is sufficient to cause injury.

### Predictive Modelling

A predictive benchmark analysis (described in appendix) was carried out to identify which method can most reliably predict whether a patient is likely to fall. For each prediction strategy, goodness of prediction is estimated by repeatedly splitting the data into a training sample on which the model is fitted and a test sample which mimicks "new" data, and on which the model is evaluated by comparing its predictions to the actual labels (faller vs non-faller).

All prediction strategies are compared on the same training/test splits, so that differences in performance can be attributed to the prediction strategy. The results of the predictive analysis are quantitative measures of how reliable each prediction strategy is in predicting new data, in terms of mean misclassification error (MMCE), sensitivity (= True Positive Rate), specificity (= True Negative Rate), precision (= Positive Predictive Value), and F1 score ( = 2 TP/(2 TP + FN + FP)).

The prediction strategies considered are different combinations of (i) the types of models used (summarized in table Y) and (ii) selected sub-sets of all variables in the data set to use in prediction. Different sets of variables are defined by using some or all of demographics, one or several of the three neurophysiological tests (Stroop, Trail, Semantic), and the Walk-12 PROM. For example, a (i) logistic regression model using (ii) demographic variables only.

Standard errors for prediction error statistics were computed by Jackknife resampling on the test folds. The performance of two strategies was considered significantly different at 5% significance level of a Wilcoxon signed-rank test conducted on the paired sample of bootstrapped (by the Jackknife) error statistics on the test folds. A strategy was considered to predict better than an uninformed guess if had a significantly lower MMCE than the uninformed predictor of always predicting "non-faller".

Receiver Operator Characteristics (ROC) of prediction strategies were computed by varying the threshold for the predictive probability of the respective methods. Bootstrap confidence bands were computed for the false positive rate (= 1 – specificity) at a 5% level of confidence.

## Ethical Considerations and Data Protection

Guidance on the nature of the study was sought from the UK Health Research Authority (HRA) who determined that the appropriate designation was 'Service Development'. The study was subsequently vetted by hospital governance and R&D groups. Patient (oral) consent to participate in the study was obtained and recorded in the clinical notes. Data analysis was conducted on a completely anonymised dataset. Non-anonymized data was stored securely for use by the patient's clinical team, accessible only through the hospitals secured severs.

# Results

An overview of all of the results obtained in our predictive analysis can be found in the appendix. Below we present a selection of these results, focusing on the four most pertinent findings.

**1. The Trail test, by itself, produces the best predictions. Moreover, addition of other neuropsychological, demographic, or physical function-related variables, to the Trail test data, doesn't improve the model.**

Table 1 presents the goodness of prediction obtained from using only the demographical variables or variables from the three neuropsychological tests, or the Walk-12. The prediction goodness is reported for the best method, among those reported in the appendix. It may be observed that the trail test makes the best predictions (Wilcoxon signed-rank $p < .001$), though it should be mentioned that each neuropsychological test is missing for a different and substantial set of patients (around 1/3), therefore differences in measures of prediction goodness may in principle arise not only from the prediction strategy but also from the different patient sample. However, adding any of the other variables does not significantly improve the goodness of prediction (Wilcoxon signed-rank residuals $p < .001$) on the subsets of patients on which such predictions are possible (see appendix).

**Table 1: Best possible prediction from the five different variable sets.**

| Dataset Utilized | Best Method | Mean Misclassification Error (MMCE) | Sensitivity | Specificity | Precision | F1 - Score |
|---|---|---|---|---|---|---|
| **Demographics** | SVM (Gauss) | 0.139 (± .019) | 0.153 (± .049) | 0.996 (± .004) | 0.833 (± .118) | 0.231 (± .075) |
| **Stroop test** | Naïve Bayes | 0.153 (± .025) | 0.508 (± .084) | 0.924 (± .020) | 0.585 (± .091) | 0.371 (± .080) |
| **Trail test** | Random Forest | **0.117** (± .022) | 0.550 (± .083) | 0.958 (± .015) | 0.758 (± .085) | **0.619** (± .071) |
| **Semantic** | LDA | 0.161 (± .023) | 0.245 (± .067) | 0.965 (± .013) | 0.604 (± .121) | 0.311 (± .080) |
| **Walk – 12** | LDA | 0.169 (± .024) | 0.100 (± .045) | 0.990 (± .007) | 0.700 (± .230) | 0.153 (± .074) |

**For each of the five variable sets (demographics and four neuropsychological tests, columns), the following are reported: the (subjectively chosen) best strategy of prediction for that variable set (second column), and measures of prediction goodness for that strategy, including Jackknife-estimated standard errors (five rightmost columns). Goodness of prediction is on the full population which has all variables predicted from available.**

**2. The best statistical strategy for predicting falls appears to be a random forest.**

Table 2 presents the goodness of prediction for selected prediction strategies on the patients which have the trail test available, hence differences are directly attributable to the method used. The results clearly demonstrate that the optimal (MMCE) values are produced by the random forest method, when compared with the logistic regression or majority prediction (Wilcoxon signed-rank on residuals $p < .001$). Moreover, the trail test appears superior to simple demographic data, such as age, which are well documented risk factors for falls across both model types (Wilcoxon signed-rank on residuals $p < .001$). Figure 1 compares the receiver operating characteristics (ROC) for the two Logistic Regression baselines as well as the Random Forest method.

**Table 2: Comparison of selected methods on patients with available trail test.**

| Method | Data Utilized | Mean Misclassification Error (MMCE) | Sensitivity | Specificity | Precision | F1 - Score |
|---|---|---|---|---|---|---|
| **Random Forest** | Trail Test | **0.117** (± .022) | 0.550 (± .083) | 0.958 (± .015) | 0.758 (± .085) | **0.619** (± .071) |
| | Demographics | 0.166 (± .026) | 0.200 (± .068) | 0.976 (± .012) | 0.667 (± .148) | 0.293 (± .089) |
| **Logistic Regression** | Trail Test | 0.150 (± .025) | 0.400 (± .081) | 0.953 (± .016) | 0.675 (± .104) | 0.487 (± .081) |
| | Demographics | 0.199 (± .028) | 0.033 (± .027) | 0.976 (± .012) | 0.167 (± .223) | 0.040 (± .047) |
| **Majority** | | 0.184 (± .027) | 0.000 (± .000) | 1.000 (± .000) | - | 0.000 (± .000) |

**The rows are different prediction strategies, determined by which method is used (first column), and which variables are predicted from (second column. Measures of prediction goodness for that strategy, including Jackknife-estimated standard errors (five rightmost columns). Goodness of prediction is on the population which has the trail test available. Note that all patients with trail test available also have the demographics variables available.**

**Figure 1 – The Receiver Operating Characteristics (ROC) for Random Forest and Logistic Regression based classifiers**

The data sets upon which the following ROCs are based was the restricted data set consisting of those with trail data (excluding those for which the trail data was missing). The figure illustrates the second conclusion that the random forest (RF) based predictor appears to be superior to that of logistic regression (LogReg) when both utilize only the Trails data. Moreover, both of these models are superior to the baseline model of demographic (Demog) data (consisting of common risk factors for falls) and the logistic regression, which suggests that direct cognitive and neuropsychological measurement (Trail) appear to improve predictive capabilities, at least in our dataset. The Area under the ROCs (AUROCs) are LogReg on Demog (0.65), LogReg on Trail (0.78), and RF on Trail (0.87).

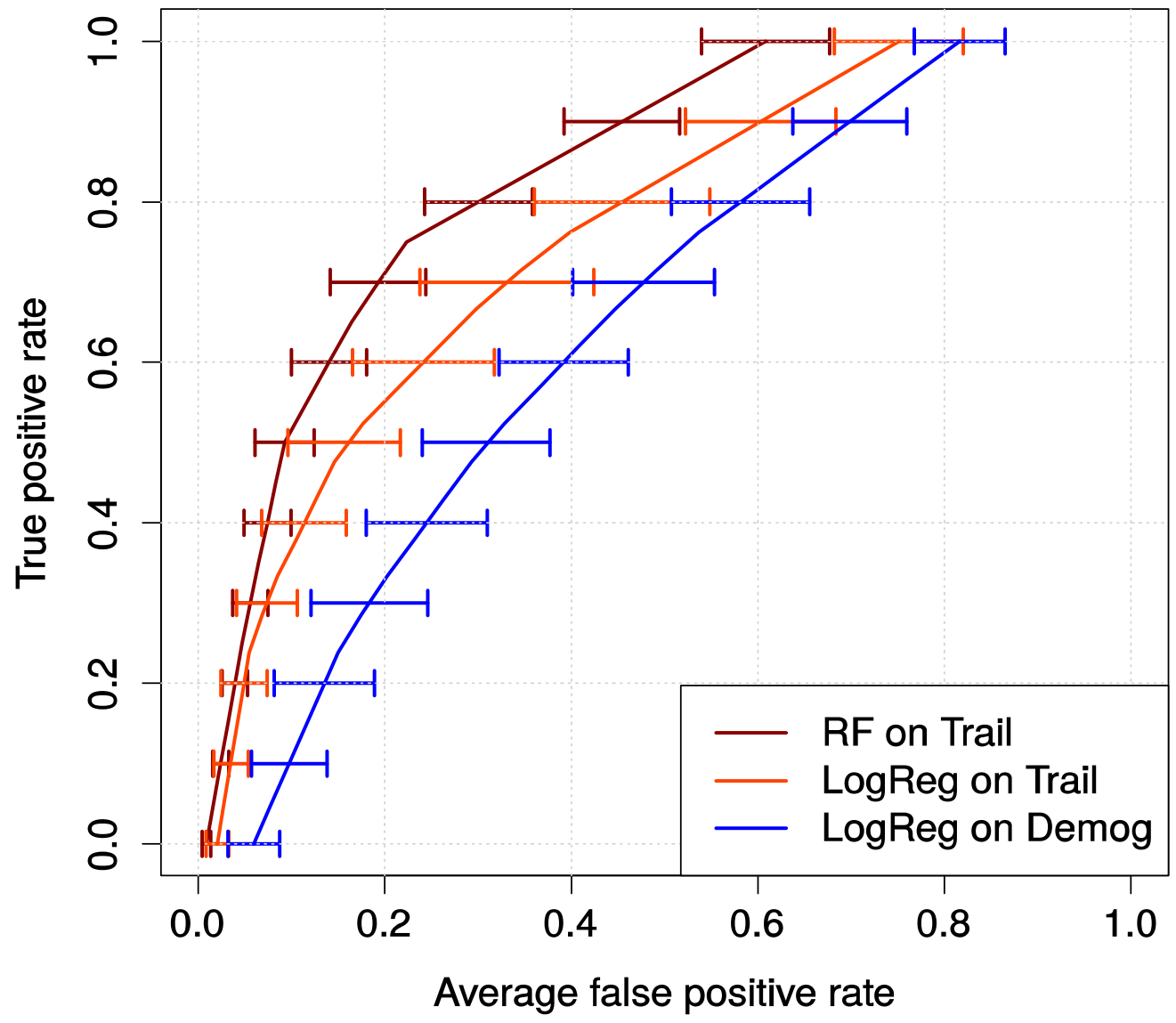

**3. The random forest strategy on the trail test is still the best method to predict falls after accounting for missing data.**

The above paragraphs show that random forest on the trail test can reliably predict falls on the patients who have a trail test recorded. Since the fact that a trail test recorded may introduce a bias, it needs to be checked how the predictive approach generalizes to the whole population. For this, prediction strategies which would use a trail test are replaced by the majority prediction (no fall) whenever the trail test is not available. It can be observed that the Random Forest utilizing Trail test data is still excellent in this real world setting, predicting better (Wilcoxon signed-rank on residuals $p < .001$) than the other methods considered.

**Table 3: Comparison of selected methods on the full patient population.**

| Prediction Model | Mean Misclassification Error (MMCE) | Sensitivity | Specificity | Precision | F1 - Score |
|---|---|---|---|---|---|
| Random Forest Tuned (Trail Test Population) & Majority (Remaining Population) | **0.119** (± .018) | 0.409 (± .067) | 0.975 (± .009) | 0.758 (± .085) | **0.508** (± .069) |
| Random Forest (Demographic Data) | 0.157 (± .020) | 0.080 (± .036) | 0.990 (± .006) | 0.667 (± .218) | 0.125 (± .061) |
| Logistic Regression (Trail Test Population) & Majority (Remaining Population) | 0.140 (± .019) | 0.294 (± .062) | 0.972 (± .010) | 0.675 (± .104) | 0.392 (± .072) |
| Logistic Regression (Demographic Data) | 0.160 (± .020) | 0.000 (± .000) | 1.000 (± .000) | - | 0.000 (± .000) |
| Majority | 0.160 (± .020) | 0.000 (± .000) | 1.000 (± .000) | - | 0.000 (± .000) |

**The rows are different prediction strategies, determined by which method is used (first column), and which variables are predicted from (second column. Measures of prediction goodness for that strategy, including Jackknife-estimated standard errors (five rightmost columns). Goodness of prediction is on the full population. Whenever a trail test is not available for a prediction strategy that otherwise uses the trail test, a majority prediction is performed.**

**4. Altering the predictive threshold allows us to produce a highly specific and sensitive tool**

Threshold tuning allows us to trade off specificity for sensitivity of the random forest model. For example, on the trail sub-group, we are able to achieve 68% (± 7.7) sensitivity, 90% (± 2.3) specificity, 0.600 (± 7.6) precision, and 0.630 (± 0.063) F1-score. On the whole population where trail test data can be missing one obtains the following values: 51% (± 6.9) sensitivity, 94% (± 1.4) specificity, 0.600 (± 0.076) precision, and 0.533 (± .062) F1-score. When overall accuracy is the desired outcome, the model described here is capable of a maximum of 76% precision (PPV), with only a modest reduction in sensitivity (Table 2 and 3).

# Discussion

This study describes what we believe is the first attempt to develop a tool for falls prediction that uses cognitive variables in a predictive modelling context. The predictive paradigm, coupled with our unique data set, has identified a *single cognitive test* – the trail test – as the most informative predictor for falling in a neurological population with an unprecedented degree of accuracy, sensitivity and specificity. This is in stark contrast to previous models including a large number of variables and lower predictive power, and suggests that our initial hypothesis regarding the interpretability and reproducibility have been a justified concern.

### Falls Prediction Tools in Context

The current gold standard measure in falls prediction is the STRATIFY questionnaire[13,21]. There are two further, more recent tools to predict falls in a neurological population: Yoo et al. use a combination of risk factors (including assessment of gait, and insight into gait ability) coupled with logistic regression analysis[22]; Kabeshova et al., have identified a neural network that is able to predict falling[18]. A summary of these three methods is displayed in Table 4.

**Table 4: Comparison with state-of-art methods on acute neurological populations.**

| Name/Method | Prediction Strategy | Sens. | Spec. | PPV | Predictive | Interpretable | Useable | Reference |
|---|---|---|---|---|---|---|---|---|
| **Proposed method** | Random Forest on Trail Test | 68% (± 7.7%) | 90% (± 2.3%) | 60% (± 7.6) | Yes | Yes | Yes | Above |
| **STRATIFY** | Logistic regression on Risk Factors data | 67% | 57% | 9.3% (Yoo et al) | No | No | Yes | [14] |
| **Yoo et al., 2015** | Logistic regression on Prospective Risk Factors data | 84.4% | 86% | 16.4% | No | Yes | No | [21] *Note: The fall rate was 3%* |
| **Kabeshova et al., 2016** | NEAT on Prospective Risk Factors data | 48.2 % | 88.0% | 65.1% | Yes | No | No | [18] |

**Columns are, from left to right: the name of method; how it predicts: sensitivity/specificity/PPV as reported in the manuscript, hence some are without confidence intervals as they have not been predictively evaluated. A question mark "?" is used when the statistic in question was not reported. Next, three questions are proposed: whether predictive validation was employed for the method; whether the prediction is clinically/neurologically interpretable; and finally, whether applying the method in clinical practice is easy and straightforward based on the infrastructure provided.**

Our approach improves on the state-of-the-art, as reported above, in a number of ways:

**Accuracy in a clinical context.** Methods with low specifity are not useful in a clinical neurological setting where the majority of patients do not fall- since a low specificity model in practice will lead to wasted resources (e.g. staff time, cost of any interventions), and even may induce a fear of falling, which can be more detrimental than actually falling with regards to health-related quality of life[23]. Further, for the purpose of directing scarce resources, the positive predictive value (PPV) of the other strategies is quite low. Our model is capable of a maximum PPV of 76% (Table 2 and 3). Hence, in terms of accuracy, we are able to offer the best tool available.

**Usability in a clinical context.** Our prediction is based solely on the trail test which can be easily conducted without the patients even leaving their beds. Hence it can be applied in everyday practice similar to the STRATIFY tool, while the other approaches would require a larger number of variables measured.

**Predictive reliability.** Predictive error estimation guarantees that the performance will not degrade, as long as the strategy is applied to a similar population – unlike for example the STRATIFY tool or non-predictive logistic regression where this has been observed before. Of course it remains to be seen whether it is the models or only the predictive modelling strategy that generalizes well to new types of populations.

**Scientific parsimony.** Our predictive model is parsimonious and well-interpretable: the fact that the trail test in itself allows the best prediction is scientifically interesting and points towards a number of hypotheses that may lead to novel insights on the interaction of human cognitive abilities and falling, or more generally overall risk assessment in a neurological population.

### Strengths, Weaknesses and Further Research

One of the main limitations of this study is a result of the data being collected in a single tertiary centre that covered acute neurological, neurosurgical, and neurorehabilitation care, suggesting that the generalization of these results should be considered carefully. Specifically, the neurorehabilitation population of patients represented less than 10% of the total, and faller cohorts, and therefore, we would argue that if this tool is utilized by other hospitals which may have a different composition of patients, the predictive accuracy on the particular population needs to be checked first – this can be done easily by running the code we have provided in appendix F and checking the estimated statistics of prediction goodness on new data. A subtler question to answer is whether the final models transfer easily between population, or whether it is the best strategy (e.g., trail test & random forest) which can yield a different model for each hospital.

Efficacy in a dedicated neurorehabilitation unit remains to be demonstrated, and this represents an important avenue for future research.

The main strength of this study is that our model is based on direct measurements of cognition, rather than proxy measurements, which are likely to be more affected by confounding factors. Moreover, the information collected in our study is already collected in the course of a clinical work-up for many patients in the in-patient neurological setting. For example, the Trail Making test is part of the standard neuropsychological evaluation at the hospital in which this study was conducted, and is widely recognized as being useful in stroke patients for a number of reasons[24]. The burdens associated with data collection are greatly reduced in such a situation, because it allows for recycling of information that is already generated for other uses.

**Implications for Policy Makers, and Clinicians**

The primary implication of this study is that a novel, highly sensitive and specific tool, for predicting falls in the acute neurological population, which surpasses the capabilities of the other tools available in this setting, is now available to policy makers and clinicians. However, we argue that the results presented here are important, not only because of their predictive power, but also because they demonstrate the efficacy of the two paradigm shifts we described earlier. In future studies, prediction models should be focus more acutely on the theoretical relevance of the data collected, with regards to the outcome being predicted, as was originally highlighted as being important. It is only then, that the true power of these modern statistical techniques will be fully realized.

## Conclusion

Although we must be cautious in making any definitive conclusions, it seems reasonable to suggest machine learning could improve the predictive faculties of future generations of predictive tools. Furthermore, the highly notable predictive power associated with the use of direct measurements of cognitive function highlights an important avenue for future research in falls risk prediction.

# Supplementary Appendix

**Table of Contents**

## Investigators

Mr. B.A. Mateen

Mr. M. Bussas

Dr. C. Doogan

Dr. D. Waller

Dr. A. Saverino

Dr. F. J. Király

Prof. E. D. Playford

# Cohort, Demographics, inclusion/exclusion criteria, and statistical analysis

### Patient Recruitment

Data was collected between the 17th November 2014 and 17th December 2014 at the National Hospital for Neurology & Neurosurgery, a tertiary neuroscience centre from a prospective cohort of 323 patients from 3 neurosurgical, 3 neurological, and 2 neuro-rehabilitation wards.

The exclusion criteria for the study included: non-fluency of English, inability to provide informed consent because of severe cognitive impairment, communication difficulties, severe mood or behavioral problems, and specific contra-indication for each test that have been highlighted in appendix Table 1.

### Demographics

The mean time from admission to testing was 4.46 days (s.d. 8.66) for the prospective cohort. The demographics for the fallers and non-fallers were then compared using two-tailed t-tests. The p-value, illustrating the degree of significance in the difference between fallers and non-fallers, has been reported. Age, number of years of formal education and ethnicity did not significantly differ between the faller and non-faller cohorts. However, there were significantly more men ($p<0.05$) in the non-faller cohort and the vast majority of both groups identified as white ethnicity (Table S3).

### Statistical Analysis

The scores on all three neuropsychological tests were described using 6 number summaries (minimum, 1st and 3rd quartile, median, mean, and maximum values). The mean of the sets for the fallers and non-fallers were then compared using two-sided t-tests. The p-value, illustrating the degree of significance in the difference between fallers and non-fallers mean score, has been reported. The error scores, composite scores, etc. were all analyzed similarly.

# Method - Predictive Benchmark Analysis

A predictive benchmark analysis was carried out to find a method which can reliably predict whether a patient is likely to fall. The result of the predictive analysis, for each method, is an estimate of how reliable the method is in predicting on new data (as opposed to classical, descriptive analysis which estimate how well a model fits existing data).

In a predictive benchmark analysis, a number of prediction strategies are compared. Those *prediction strategies* are specified by the following:

(a) Which variable is *predicted* (the so-called target variable or target outcome). Here, this is always whether the patient has fallen or not.

(b) Which variables the *prediction is based on* (the so-called covariates or features). Here, the selection is made from among the three different neuropsychological tests, a PROM of physical function, and demographic data, as detailed in [table S4 - 8].

(c) Which statistical or machine learning *method* is used for prediction. Table S1 contains an overview over the different methods used. The methods considered may be roughly divided in "classical" models such as logistic regression, and "machine learning" methods such as random forests, though this distinction is more historical than principled. The majority predictor, which always predicts that a patient did not fall, plays an important role: it is added as a "stupid"/uninformed baseline, since only in comparison to such a baseline one can say that the other method is better than a random guess.

**Implementation**

The experiments were performed using the *R (v 3.2.0)* statistical software suite and the *mlr (v 2.7)* machine learning library. Full code from our analysis can be downloaded from the attached link [to be added after publication]. Data set is available in a censored and variable-reduced form (to prevent de-anonymization by diagnosis code) via the same link [to be added after publication]. A full version is available upon request.

**Table S1. Overview of prediction methods used in the benchmark analysis.**

| Predictive Strategy | Purpose for utilizing the strategy | Prediction rationale | Literature references for specific technical details |
|---|---|---|---|
| Logistic Regression | Statistical models often considered "classical". Logistic regression would be the standard explanatory/descriptive strategy for modelling falls in our setting. | Standard logistic regression models the log-odds of falling as a linear function in the selected covariates. | (Hastie et al, section 4.4)[1] (Cox, 1958)[2] |
| Linear Discriminant Analysis (LDA) | | LDA attempts to separate the two classes of fallers and non-fallers by a linear functional which minimizes in-class variance while maximizing between-class variance. | (Hastie et al, section 4.3)[1] (Fisher, 1936)[3] |
| Naïve Bayes | | Naïve Bayes classifiers probabilistically predict falling/non-falling from the covariates based on Bayes' theorem, under the simplifying but possibly inaccurate assumption that the covariates are independent. | (Hastie et al, section 6.6.3)[1] (Maron & Kuhns 1960)[4] |
| Kernel Support Vector Machine (soft-margin C-SVC with linear kernel or Gaussian kernel) | Three modern machine learning methods considered to be some of the best general-purpose classifiers that exist. | Support vector classifiers construct a hyperplane to separate (most) fallers and non-fallers with maximum margin. We employ the frequently used kernel variant which allows for a non-linear separating hyperplane. | (Hastie et al, section 12)[1] (Cortes & Vapnik, 1995)[5] |
| Random Forests | | The random forest classifier constructs a large number of decision trees based on the covariates to predict whether the patient falls. The decision trees are aggregated ("bagged") to overcome the known predictive deficiency ("overfitting") of a single decision tree classifier. | (Hastie et al, section 15)[1] (Breiman, 2001)[6] |
| Neural Networks | | Neural network classifiers construct a complex classificiation function with a particular substitution structure inspired by real world neural networks. The function is fitted to the data by gradient descent; the structure of the network varies greatly in literature, we use the two standard variants in mlr (nnet, avvnet) | (Hastie et al, section 11)[1] (Werbos, 1974)[7] (Parker, 1985)[8] |
| Majority Prediction | A "simple guess" for later quantitative comparisons "better than a simple guess" | Predicts that no patient is going to fall. | N/A |

## Validation set-up

In order to assess how well the prediction strategies, predict fallers on new, unseen data, a validation experiment in performed which mimics exactly that process: each strategy is used to fit parameters of the method on part of the data, the so-called training data. Prediction is then performed on other part of the data, the test data, which plays the role of the new data. Goodness of prediction is evaluated by comparing the prediction (here: whether the patient falls) to the true target variable on the test data. All methods are compared on the same training/test splits, hence significant differences may be attributed to the method.

This way of evaluation is standard to estimate the predictive goodness of a prediction strategy and is called *out-of-sample validation*, predictive model evaluation, or independent test set validation. Several set-ups for validation experiments which guarantee accurate estimation of the prediction error are well-known (see Hastie section 7)[1]. Our *validation set-up* is specified by the following:

(a) How the *training/test splits* are selected: 10-fold cross-validation, on the population of patients that have answered *all* questions in the neuropsychological tests that were used. For example, if the prediction is made from trail test variables only, the validation is done on all patients who have a full trail test.

(b) How *goodness of prediction is measured*. It is standard to uses multiple performance measures to assess the quality of our results. The measures of predictive goodness reported are: the mean misclassification error (MMCE), sensitivity, specificity, precision and F1-Score (harmonic mean of sensitivity and precision). Falling is considered as the positive level of the target variable.

(c) How *goodness of prediction is compared* between methods. Samples of prediction errors may be obtained from the Jackknife samples of error statistics, in each of the ten test samples. These are union-aggregated and used to obtain error bars/confidence intervals for error measures by the Jackknife estimator of variance. The samples are naturally paired between methods predicting from the same sample of patients, therefore they may be used to obtain non-parametric significances of whether one method is better than a second. Note that this method of error estimation may be biased and underestimate errors response variances of the error statistics.

## Pre-processing and tuning

Prior to prediction, all variables were normalized to have zero mean and standard deviation one on the *training set*. Whenever an advanced prediction strategy required tuning of parameters, this was done by 3-fold cross-validation grid-tuning *inside the training set*. See figure S1 for a list of the tuning grids employed. Note that all pre-processing and tuning has to take place on the training set only, to correctly mimic the process of predicting on new data. Otherwise information on the new data would be implicitly used already in a phase where such data has not been yet seen.

**<u>Figure – S1: Tuning Grids</u>**

```
Linear SVM:
ps = makeParamSet(
makeDiscreteParam("C", values = 2^(-4:4))         #
C parameter of the SVM
)

Gaussian SVM
ps = makeParamSet(
makeDiscreteParam("C", values = 2^(-
2:2)),          # C parameter of the SVM
makeDiscreteParam("sigma", values = 2^(-
2:2))     # Bandwidth of the Gaussian kernel
)

Random Forest
ps = makeParamSet(
makeDiscreteParam("ntree", values =
c(100,250,500,1000,2000))  # Number of the trees
used in the forest
)
```

**Table S2 – Descriptions of the Three Neuropsychological Testing Paradigms, the Physical Function PROM, and Resulting Variables**

| Measure | Description (Adapted from Zomeren & Spikman, 2003)[9] | Variables | | Cognitive Processes |
|---|---|---|---|---|
| **Stroop Colour-Word Test**<br><br>**Original Citation –**<br>Stroop, 1935[10]<br><br>**Variant Utilized –**<br>Trenerry et al., 1989[11] | **Part 1 – Word Reading**<br><br>Participants are presented with the first of two cards, and are asked to read the words on the page. The stimulus on the first card consists of 112 words in three columns. The word can be any of the following four colours: Blue, Green, Red or Tan, and is printed in a colour of ink that does not correspond to the word.<br><br>**Part 2 – Colour Naming**<br><br>The second part of the test is referred to as the interference task, and in this instance the participant is asked to name the colour of ink in which the word is printed. An identical card to that which was presented in the first part is presented to the participant with the new instructions. | **Raw –**<br>1. Words read in 1 & 2 minutes<br>2. Errors on word reading<br><br>**Calculated – None**<br><br>**Raw –**<br>1. Colours named in 1 & 2 minutes<br>2. Errors on colour naming<br>3. Colour naming error corrections<br><br>**Calculated –**<br>4. Proportion of Errors Corrected<br>5. Number of colours named in 2 minutes divided by the number of words reads in 2 minutes | **Specific Contraindications –**<br>If the participant needs visual aids to read, ensure that they are used during testing.<br>If the participant is too visually impaired or has a condition, such as colour blindness, abandon testing.<br><br>**Notes –**<br>The maximum score is 112 | **Part 1 – Word Reading**<br>(Error scores not relevant to stated cognitive process)<br><br><u>Attentional Process:</u> An operational level task measuring the speed of information processing.<br><br>**Part 2 – Colour Naming**<br>(Error scores related to stated cognitive process)<br><br><u>Executive Functions:</u> Response Inhibition<br><br><u>Attentional Process:</u> A tactical level task measuring focused attention |
| **Trail Making**<br><br>**Original Citation -**<br>Army Individual Test Battery, 1944[12]<br><br>**Variant Used –**<br>Reitan, 1986[13] | **Part A – Number Task**<br><br>The participant is presented with the stimulus and asked to draw a line joining consecutively numbered circles from 1 -25, as quickly as they can.<br><br>**Part B – Number/Letter Task**<br><br>The participant is presented with the second stimulus and asked to draw a line joining consecutively numbered (1-13) and lettered circles (A-L), by alternating between the two types of sequences. | **Raw –**<br>1. Time taken to complete number task<br>2. Errors on number task<br><br>**Calculated - None**<br><br>**Raw –**<br>1. Time taken to complete number/letter task<br>2. Errors on number/letter task<br><br>**Calculated –**<br>3. Time to complete number/letter task divided by the time to complete number task | **Specific Contraindications –**<br>The test must be completed using the participant's dominant arm therefore hemi-paresis on the dominant side is a contra-indication. Severe visual deficits is also a contra-indication<br><br>**Notes –**<br>The test is to be abandoned if incomplete after 300 seconds. And errors in this situation are recorded as an unknown | **Part A – Number Task**<br>(Error scores not relevant to stated cognitive process)<br><br>Attentional Process: An operational level task measuring the speed of information processing<br><br>**Part B – Number/Letter Task**<br>((Error scores related to stated cognitive process)<br><br>Executive Function: Fluid Intelligence/Multitasking<br><br>Attentional Process: A tactical level task measuring focused attention |
| **Semantic Fluency**<br><br>**Original Citation –**<br>Thurstone, 1938[14]<br><br>**Variant Used –**<br>Described in Strauss et al., 2006[15] | Participants are asked to name as many animals as they can in a minute. Participants are told they can use any letter of the alphabet and do not need to go in any particular order. | **Raw –**<br>1. Number of animals<br>2. Number of repetitions | **Specific Contraindications –**<br>For certain severely aphasic or dysphasic individuals, circumstances should be evaluated to determine whether this test is appropriate. | **Number of Animals -**<br>Executive Function:<br><br>Attentional Process: A tactical level task measuring focused attention & an operational level task measuring the speed of information processing.<br><br>**Error Scores** –<br>Related to certain diagnoses, such as dementias (Straus et al., 2006)[15] |
| **Walk – 12**<br>Hobart et al., 2003[16] | 12 questions with a 5 point Likert scale regarding walking, and walking-related ability. | **Raw –**<br>1. The individual score on each question<br><br>**Calculated –**<br>2. The summed score for the entire questionnaire | **Specific Contraindications –**<br>For certain severely aphasic or dysphasic individuals, circumstances should be evaluated to determine whether this test is appropriate. | **NA** |

## Table S3 – Cohort Demographics

| *Demographic Data* | *Fallers N = 54* | *Non-Fallers N = 284* |
|---|---|---|
| **Sex** | | |
| Male | 46.3% | 59.7 % |
| Female | 53.7% | 40.3 % |
| **Ethnicity*** | | |
| White | 77.8% | 77.4% |
| Asian | 9.26% | 14.2% |
| Black | 7.41% | 4.60% |
| Afro-Caribbean | 3.70% | 2.09% |
| Mixed | 1.85% | 1.67% |
| **Age** | | |
| <19 | 0.00% | 0.35% |
| 19 - 29 | 11.1% | 8.45% |
| 29 - 39 | 5.56% | 15.5% |
| 39 - 49 | 18.5% | 14.8% |
| 49 - 59 | 25.9% | 20.4% |
| 59 - 69 | 18.5% | 17.3% |
| 69 - 79 | 11.1% | 16.5% |
| 79 - 89 | 9.26% | 6.34% |
| 89 – 99 | 0.00% | 0.35% |
| Mean [95% Confidence Interval] | 55.4 [50.7, 60.2] | 54.7 [52.5, 56.9] |
| **Years of Education^** | | |
| Mean [95% Confidence Interval] | 13.1 [12.1, 14.1] | 13.4 [12.9, 14.0] |

| **Diagnoses** |
|---|
| 1. Undefined; 2. Achondroplasia; 3. Ankylosing Spondylitis; 4. Autonomic Diseases & Disorders; 5. Brain Tumor – Frontal; 6. Brain Tumor – Cerebellar; 7. Brain Tumor – Occipital; 8. Brain Tumor – Other; 9. Brain Tumor – Parietal; 10. Brain Tumor – Temporal; 11. Cauda Equina Syndrome; 12. Central Cord Syndrome & Syringomyelia; 13. Cerebral Palsy; 14. Chiari Malformation; 15. Chronic Fatigue Syndrome; 16. Cognitive Decline; 17. Cushing's Disease & Syndrome; 18. Depressive Disorders; 19 Dropped Head Syndrome; 20. Drug Abuse; 21. Dystonia – Focal; 22. Epilepsy; 23. Encephalopathy; 24. Functional Movement Disorder; 25. Guillian Barre Syndrome; 26. Headache & Migraine; 27. Hydrocephalus; Intrancranial Hypertension; 29. Motor Neuron Disease; 30. Multiple Sclerosis; 31. Myasthenia Gravis; 32. Myelopathy – Cervical; 33. Myelopathy – Lumbar; 34. Myelopathy – Other; 35. Myelopathy – Thoracic; 36. Myopathy; 37. Neuropathy; 38. Other Neurovascular Disorder/Disease; 39. Parkinson's & Parkinson's-like Disorders; 40. Phenylkentonuria; 41. Foot Drop; 42. Schizophrenia; 43. Spina Bifida; 44. Spinal Stenosis; 45. Spinal Tumor – Lumbar; 46. Spinal Tumor – Thoracic; 47. Stiff Person Syndrome; 48. Stroke; 49. SUNA & SUNT; 50. Tremor – Dystonic; 51. Tuberculosis; 52. Tumor – Other. |

*Ethnicity reported in line with the standardized classification used by the office for national statistics[17].
^ Total number of years in primary, secondary, further &/or higher education.

## Summary Statistics for Raw Data

*Note: significances in this appendix are not post-hoc/multiple testing corrected*

**Table S4 – Non-Faller and Faller Summary Statistics, Discrete variables.** **Last column is Fallers vs non-Fallers, as measured by Pearson's Chi-squared test.**

| Test | Population | Sample Size (Participants) | Yes | No | Significance (Chi-squaredtest) |
|---|---|---|---|---|---|
| Theatre in the last month? | Faller | 54 | 31 | 23 | $4.6 \times 10^{-1}$ |
| | Non-Faller | 283 | 144 | 139 | |
| Fallen in the last month? | Faller | 54 | 29 | 25 | $4.6 \times 10^{-4}$ |
| | Non-Faller | 283 | 80 | 203 | |
| Physical function change in the last month? | Faller | 54 | 43 | 11 | $1.8 \times 10^{-3}$ |
| | Non-Faller | 283 | 158 | 125 | |

**Table S5 -Non-Faller and Faller Summary Statistics, Trail test.** **Last column is Fallers vs non-Faller means, as measured by Student's t-test.**

| Test | Population | Sample Size (Participants) | Minimum | 1st Quartile | Median | Mean | 3rd Quartile | Maximum | Significance (t-test) |
|---|---|---|---|---|---|---|---|---|---|
| Time to Complete Part A (Seconds) | Faller | 39 | 15.0 | 49.0 | 76.0 | 80.43 | 90.5 | 300.0 | $3.0 \times 10^{-5}$ |
| | Non-Faller | 172 | 14.0 | 26.0 | 34.0 | 42.51 | 48.0 | 131.0 | |
| Number of Errors - Part A | Faller | 38 | 0.0 | 0.0 | 0.0 | 1.10 | 1.0 | 3.0 | $1.2 \times 10^{-2}$ |
| | Non-Faller | 172 | 0.0 | 0.0 | 0.0 | 0.93 | 0.0 | 2.0 | |
| Time to complete Part B (Seconds) | Faller | 39 | 42.0 | 176.0 | 253.0 | 200.71 | 294.5 | 300.0 | $3.9 \times 10^{-8}$ |
| | Non-Faller | 171 | 32.0 | 84.0 | 131.0 | 121.57 | 191.0 | 300.0 | |
| Number of Errors - Part B | Faller | 38 | 0.0 | 0.3 | 2.0 | 1.27 | 3.0 | 8.0 | $2.0 \times 10^{-4}$ |
| | Non-Faller | 168 | 0.0 | 0.0 | 0.0 | 0.80 | 1.0 | 7.0 | |
| Time to Complete Part B / Time to Complete Part A | Faller | 38 | 1.0 | 2.4 | 2.8 | 2.79 | 4.2 | 10.5 | $4.3 \times 10^{-1}$ |
| | Non-Faller | 171 | 1.6 | 2.5 | 3.6 | 2.97 | 4.6 | 7.9 | |

**Table S6 – Non-Faller and Faller Summary Statistics, Stroop Colour-Word Tests.** **Last column is Fallers vs non-Faller means, as measured by Student's t-test.**

| Test | Population | Sample Size (Participants) | Minimum | 1st Quartile | Median | Mean | 3rd Quartile | Maximum | Significance (t-test) |
|---|---|---|---|---|---|---|---|---|---|
| Number of Words Read in 1 Minute - Part A | Faller | 37 | 20.0 | 49.0 | 66.0 | 78.15 | 104.0 | 112.0 | $7.7 \times 10^{-7}$ |
| | Non-Faller | 178 | 21.0 | 93.0 | 112.0 | 98.23 | 112.0 | 112.0 | |
| Number of Words Read in 2 Minutes - Part A | Faller | 37 | 38.0 | 100.0 | 112.0 | 103.42 | 112.0 | 112.0 | $4.7 \times 10^{-3}$ |
| | Non-Faller | 180 | 46.0 | 112.0 | 112.0 | 111.27 | 112.0 | 112.0 | |
| Number of Errors - Part A | Faller | 37 | 0.0 | 0.0 | 0.0 | 0.85 | 1.0 | 7.0 | $1.2 \times 10^{-1}$ |
| | Non-Faller | 180 | 0.0 | 0.0 | 0.0 | 0.99 | 1.0 | 6.0 | |
| Number of Corrected Errors - Part A | Faller | 37 | 0.0 | 0.0 | 0.0 | 0.42 | 1.0 | 2.0 | $5.6 \times 10^{-1}$ |
| | Non-Faller | 180 | 0.0 | 0.0 | 0.0 | 0.35 | 0.0 | 4.0 | |
| Number of Colours Identified in 1 Minute - Part B | Faller | 37 | 4.0 | 26.0 | 36.0 | 38.00 | 51.0 | 70.0 | $1.3 \times 10^{-4}$ |
| | Non-Faller | 174 | 20.0 | 42.0 | 50.0 | 47.27 | 59.8 | 112.0 | |
| Number of Colours Identified in 2 Minutes - Part B | Faller | 37 | 9.0 | 54.0 | 72.0 | 73.78 | 100.0 | 112.0 | $1.4 \times 10^{-4}$ |
| | Non-Faller | 174 | 35.0 | 82.3 | 100.0 | 88.84 | 112.0 | 112.0 | |
| Number of Errors - Part B | Faller | 37 | 0.0 | 2.0 | 3.0 | 3.91 | 5.0 | 12.0 | $5.2 \times 10^{-4}$ |
| | Non-Faller | 174 | 0.0 | 0.0 | 1.0 | 2.45 | 3.0 | 10.0 | |
| Number of Corrected Errors - Part B | Faller | 37 | 0.0 | 1.0 | 2.0 | 2.31 | 3.0 | 7.0 | $3.7 \times 10^{-2}$ |
| | Non-Faller | 174 | 0.0 | 0.0 | 1.0 | 1.70 | 2.0 | 8.0 | |

**Table S7 – Non-Faller and Faller Summary, Semantic Fluency Tests.** **Last column is Fallers vs non-Faller means, as measured by Student's t-test.**

| Test | Population | Sample Size (Participants) | Minimum | 1st Quartile | Median | Mean | 3rd Quartile | Maximum | Significance (t-test) |
|---|---|---|---|---|---|---|---|---|---|
| Number of Animals | Faller | 44 | 5.0 | 10.0 | 14.0 | 15.70 | 18.0 | 29.0 | $8.5 \times 10^{-7}$ |
| | Non-Faller | 209 | 3.0 | 15.0 | 19.0 | 19.77 | 25.0 | 38.0 | |
| Number of Repetitions | Faller | 44 | 0.0 | 0.0 | 1.0 | 0.91 | 2.0 | 4.0 | $4.6 \times 10^{-3}$ |
| | Non-Faller | 205 | 0.0 | 0.0 | 0.0 | 0.64 | 1.0 | 3.0 | |

**Table S8 – Non-Faller and Faller Summary, Walk-12 Test.** **Last column is Fallers vs non-Faller means, as measured by Student's t-test.**

| Test | Population | Sample Size (Participants) | Minimum | 1st Quartile | Median | Mean | 3rd Quartile | Maximum | Significance (t-test) |
|---|---|---|---|---|---|---|---|---|---|
| Question 1 | Faller | 50 | 0.0 | 2.3 | 4.0 | 3.88 | 5.0 | 5.0 | $9.5 \times 10^{-3}$ |
| | Non-Faller | 204 | 0.0 | 1.0 | 3.0 | 2.91 | 4.0 | 5.0 | |
| Question 2 | Faller | 50 | 0.0 | 2.0 | 5.0 | 4.28 | 5.0 | 5.0 | $1.8 \times 10^{-1}$ |
| | Non-Faller | 204 | 0.0 | 1.0 | 4.0 | 3.34 | 5.0 | 5.0 | |
| Question 3 | Faller | 50 | 0.0 | 2.0 | 4.0 | 3.81 | 5.0 | 5.0 | $2.3 \times 10^{-2}$ |
| | Non-Faller | 204 | 0.0 | 1.0 | 3.0 | 2.89 | 4.0 | 5.0 | |
| Question 4 | Faller | 50 | 0.0 | 2.0 | 4.0 | 3.53 | 5.0 | 5.0 | $1.2 \times 10^{-1}$ |
| | Non-Faller | 204 | 0.0 | 1.0 | 3.0 | 2.95 | 4.0 | 5.0 | |
| Question 5 | Faller | 50 | 0.0 | 2.0 | 3.5 | 3.53 | 5.0 | 5.0 | $2.6 \times 10^{-1}$ |
| | Non-Faller | 204 | 0.0 | 1.8 | 3.0 | 2.87 | 4.0 | 5.0 | |
| Question 6 | Faller | 50 | 0.0 | 3.0 | 4.5 | 4.19 | 5.0 | 5.0 | $6.7 \times 10^{-2}$ |
| | Non-Faller | 204 | 0.0 | 2.0 | 3.0 | 3.29 | 5.0 | 5.0 | |
| Question 7 | Faller | 50 | 0.0 | 2.3 | 4.5 | 4.13 | 5.0 | 5.0 | $4.9 \times 10^{-2}$ |
| | Non-Faller | 204 | 0.0 | 2.0 | 3.0 | 3.17 | 4.0 | 5.0 | |
| Question 8 | Faller | 50 | 0.0 | 1.0 | 4.0 | 3.72 | 5.0 | 5.0 | $1.8 \times 10^{-1}$ |
| | Non-Faller | 204 | 0.0 | 1.0 | 3.0 | 2.83 | 5.0 | 5.0 | |
| Question 9 | Faller | 50 | 0.0 | 1.0 | 4.0 | 3.91 | 5.0 | 5.0 | $6.1 \times 10^{-2}$ |
| | Non-Faller | 204 | 0.0 | 1.0 | 2.0 | 2.68 | 5.0 | 5.0 | |
| Question 10 | Faller | 50 | 0.0 | 3.0 | 5.0 | 4.13 | 5.0 | 5.0 | $6.5 \times 10^{-2}$ |
| | Non-Faller | 204 | 0.0 | 2.0 | 3.0 | 3.27 | 5.0 | 5.0 | |
| Question 11 | Faller | 50 | 0.0 | 3.0 | 4.0 | 4.09 | 5.0 | 5.0 | $6.7 \times 10^{-2}$ |
| | Non-Faller | 204 | 0.0 | 1.0 | 3.0 | 3.14 | 5.0 | 5.0 | |
| Question 12 | Faller | 50 | 0.0 | 4.0 | 5.0 | 4.44 | 5.0 | 5.0 | $8.7 \times 10^{-3}$ |
| | Non-Faller | 204 | 0.0 | 1.0 | 4.0 | 3.25 | 5.0 | 5.0 | |

## All Predictive Analysis Results (Table S9)

| Variable Utilized | Population | Method | Mean Misclassification Error (MMCE) | Sensitivity | Specificity | Precision | F1 - Score |
|---|---|---|---|---|---|---|---|
| Demographics | All | Logistic Regression | 0.160 (±0.020) | 0.000 (±0.000) | 1.000 (±0.000) | - | 0.000 (±0.000) |
| | | Linear Discriminant Analysis | 0.160 (±0.020) | 0.000 (±0.000) | 1.000 (±0.000) | - | 0.000 (±0.000) |
| | | SVM (Linear) | 0.160 (±0.020) | 0.000 (±0.000) | 1.000 (±0.000) | - | 0.000 (±0.000) |
| | | SVM (Gauss) | 0.139 (±0.019) | 0.153 (±0.049) | 0.996 (±0.004) | 0.833 (±0.118) | 0.231 (±0.075) |
| | | Random Forest | 0.157 (±0.020) | 0.080 (±0.036) | 0.990 (±0.006) | 0.667 (±0.218) | 0.125 (±0.061) |
| | | Naïve Bayes | 0.163 (±0.020) | 0.113 (±0.044) | 0.975 (±0.009) | 0.500 (±0.150) | 0.162 (±0.065) |
| | | Neural Net | 0.169 (±0.020) | 0.230 (±0.058) | 0.948 (±0.013) | 0.496 (±0.099) | 0.245 (±0.068) |
| | | avNNet | 0.160 (±0.020) | 0.190 (±0.054) | 0.965 (±0.011) | 0.450 (±0.118) | 0.249 (±0.070) |
| | Trail | Logistic Regression | 0.199 (±0.028) | 0.033 (±0.027) | 0.976 (±0.012) | 0.167 (±0.223) | 0.040 (±0.047) |
| | | Linear Discriminant Analysis | 0.189 (±0.027) | 0.083 (±0.045) | 0.976 (±0.012) | 0.458 (±0.218) | 0.109 (±0.072) |
| | | SVM (Linear) | 0.184 (±0.027) | 0.000 (±0.000) | 1.000 (±0.000) | - | 0.000 (±0.000) |
| | | SVM (Gauss) | 0.170 (±0.026) | 0.125 (±0.056) | 0.988 (±0.008) | 0.714 (±0.199) | 0.200 (±0.086) |
| | | Random Forest | 0.166 (±0.026) | 0.200 (±0.068) | 0.976 (±0.012) | 0.667 (±0.148) | 0.293 (±0.089) |
| | | Naïve Bayes | 0.180 (±0.027) | 0.333 (±0.079) | 0.928 (±0.020) | 0.517 (±0.104) | 0.371 (±0.080) |
| | | Neural Net | 0.204 (±0.028) | 0.200 (±0.068) | 0.929 (±0.020) | 0.338 (±0.115) | 0.206 (±0.079) |
| | | **avNNet** | 0.205 (±0.028) | 0.258 (±0.073) | 0.916 (±0.021) | 0.425 (±0.105) | 0.317 (±0.079) |
| **Strrop** | **Stroop** | **Logistic Regression** | 0.153 (±0.025) | 0.367 (±0.080) | 0.953 (±0.016) | 0.633 (±0.111) | 0.454 (±0.084) |
| | | **Linear Discriminant Analysis** | 0.153 (±0.025) | 0.392 (±0.082) | 0.947 (±0.017) | 0.608 (±0.106) | 0.462 (±0.082) |
| | | **SVM (Linear)** | 0.149 (±0.025) | 0.225 (±0.069) | 0.988 (±0.0008 | 0.833 (±0.140) | 0.343 (±0.092) |
| | | **SVM (Gauss)** | 0.192 (±0.027) | 0.050 (±0.038) | 0.971 (±0.013) | 0.133 (±0.199) | 0.057 (±0.063) |
| | | **Random Forest** | 0.182 (±0.027) | 0.258 (±0.072) | 0.941 (±0.018) | 0.463 (±0.121) | 0.307 (±0.083) |
| | | **Naïve Bayes** | 0.153 (±0.025) | 0.508 (±0.084) | 0.924 (±0.020) | 0.585 (±0.091) | 0.534 (±0.075) |
| | | **Neural Net** | 0.205 (±0.028) | 0.467 (±0.084) | 0.865 (±0.026) | 0.451 (±0.080) | 0.441 (±0.072) |
| | | **avNNet** | 0.163 (±0.026) | 0.450 (±0.084) | 0.924 (±0.020) | 0.560 (±0.095) | 0.475 (±0.078) |
| **Trail** | **Trail** | **Logistic Regression** | 0.150 (±0.025) | 0.400 (±0.081) | 0.953 (±0.016) | 0.675 (±0.104) | 0.487 (±0.081) |
| | | **Linear Discriminant Analysis** | 0.154 (±0.025) | 0.425 (±0.082) | 0.941 (±0.018) | 0.675 (±0.099) | 0.514 (±0.078) |
| | | **SVM (Linear)** | 0.169 (±0.026) | 0.217 (±0.068) | 0.970 (±0.013) | 0.679 (±0.146) | 0.278 (±0.087) |
| | | **SVM (Gauss)** | 0.160 (±0.026) | 0.283 (±0.075) | 0.964 (±0.014) | 0.620 (±0.123) | 0.360 (±0.086) |
| | | **Random Forest** | 0.117 (±0.022) | 0.550 (±0.083) | 0.958 (±0.015) | 0.758 (±0.085) | 0.619 (±0.071) |
| | | **Naïve Bayes** | 0.164 (±0.026) | 0.550 (±0.083) | 0.899 (±0.023) | 0.638 (±0.083) | 0.566 (±0.070) |
| | | **Neural Net** | 0.173 (±0.027) | 0.600 (±0.081) | 0.876 (±0.026) | 0.597 (±0.077) | 0.578 (±0.067) |
| | | **avNNet** | 0.169 (±0.026) | 0.442 (±0.083) | 0.917 (±0.021) | 0.654 (±0.092) | 0.498 (±0.076) |
| **Semantic** | **Semantic** | **Logistic Regression** | 0.164 (±0.024) | 0.200 (±0.062) | 0.971 (±0.012) | 0.583 (±0.135) | 0.271 (±0.080) |
| | | **Linear Discriminant Analysis** | 0.161 (±0.023) | 0.245 (±0.067) | 0.965 (±0.013) | 0.604 (±0.121) | 0.311 (±0.080) |
| | | **SVM (Linear)** | 0.176 (±0.024) | 0.000 (±0.000) | 1.000 (±0.000) | - | 0.000 (±0.000) |
| | | **SVM (Gauss)** | 0.164 (±0.024) | 0.115 (±0.049) | 0.990 (±0.007) | 0.792 (±0.199) | 0.159 (±0.077) |
| | | **Random Forest** | 0.169 (±0.024) | 0.245 (±0.067) | 0.956 (±0.014) | 0.528 (±0.117) | 0.310 (±0.078) |
| | | **Naïve Bayes** | 0.173 (±0.024) | 0.270 (±0.069) | 0.946 (±0.016) | 0.550 (±0.109) | 0.337 (±0.077) |
| | | **Neural Net** | 0.165 (±0.024) | 0.245 (±0.067) | 0.961 (±0.014) | 0.537 (±0.119) | 0.313 (±0.079) |
| | | **avNNet** | 0.161 (±0.023) | 0.265 (±0.069) | 0.960 (±0.014) | 0.656 (±0.115) | 0.347 (±0.079) |
| **Walk-12** | **Walk-12** | **Logistic Regression** | 0.182 (±0.025) | 0.100 (±0.045) | 0.975 (±0.011) | 0.583 (±0.186) | 0.153 (±0.070) |
| | | **Linear Discriminant Analysis** | 0.169 (±0.024) | 0.100 (±0.045) | 0.990 (±0.007) | 0.700 (±0.230) | 0.153 (±0.074) |
| | | **SVM (Linear)** | 0.178 (±0.025) | 0.000 (±0.000) | 1.000 (±0.000) | - | 0.000 (±0.000) |
| | | **SVM (Gauss)** | 0.178 (±0.025) | 0.000 (±0.000) | 1.000 (±0.000) | - | 0.000 (±0.000) |
| | | **Random Forest** | 0.241 (±0.028) | 0.145 (±0.054) | 0.894 (±0.022) | 0.248 (±0.063) | 0.165 (±0.062) |
| | | **Naïve Bayes** | 0.408 (±0.032) | 0.665 (±0.073) | 0.576 (±0.035) | 0.268 (±0.041) | 0.375 (±0.050) |
| | | **Neural Net** | 0.190 (±0.025) | 0.020 (±0.023) | 0.980 (±0.010) | 0.200 (±0.223) | 0.020 (±0.042) |
| | | **avNNet** | 0.224 (±0.027) | 0.100 (±0.045) | 0.924 (±0.019) | 0.130 (±0.099) | 0.107 (±0.060) |
| **All Variables** | **All Variables** | **Logistic Regression** | 0.235 (±0.037) | 0.500 (±0.104) | 0.825 (±0.037) | 0.485 (±0.091) | 0.444 (±0.085) |
| | | **Linear Discriminant Analysis** | 0.138 (±0.030) | 0.567 (±0.103) | 0.933 (±0.024) | 0.694 (±0.108) | 0.572 (±0.088) |
| | | **SVM (Linear)** | 0.130 (±0.030) | 0.417 (±0.102) | 0.980 (±0.013) | 0.875 (±0.117) | 0.503 (±0.0103) |
| | | **SVM (Gauss)** | 0.198 (±0.035) | 0.000 (±0.000) | 0.991 (±0.010) | 0.000 (±0.000) | 0.0000 (±0.000) |
| | | **Random Forest** | 0.131 (±0.030) | 0.433 (±0.103) | 0.971 (±0.016) | 0.815 (±0.118) | 0.510 (±0.099) |
| | | **Naïve Bayes** | 0.166 (±0.033) | 0.667 (±0.100) | 0.877 (±0.032) | 0.610 (±0.095) | 0.597 (±0.081) |
| | | **Neural Net** | 0.201 (±0.035) | 0.600 (±0.103) | 0.853 (±0.034) | 0.550 (±0.096) | 0.522 (±0.085) |
| | | **avNNet** | 0.147 (±0.031) | 0.700 (±0.097) | 0.894 (±0.030) | 0.607 (±0.095) | 0.614 (±0.079) |
| **Trail + Demographics** | **Trail + Demographics** | **Logistic Regression** | 0.132 (±0.024) | 0.450 (±0.083) | 0.957 (±0.015) | 0.735 (±0.093) | 0.498 (±0.076) |
| | | **Linear Discriminant Analysis** | 0.132 (±0.024) | 0.475 (±0.083) | 0.951 (±0.016) | 0.707 (±0.091) | 0.514 (±0.074) |
| | | **SVM (Linear)** | 0.165 (±0.026) | 0.225 (±0.071) | 0.969 (±0.013) | 0.628 (±0.138) | 0.264 (±0.068) |
| | | **SVM (Gauss)** | 0.204 (±0.028) | 0.0000 (±0.000) | 0.976 (±0.012) | 0.0000 (±0.000) | 0.0000 (±0.000) |
| | | **Random Forest** | 0.136 (±0.024) | 0.450 (±0.083) | 0.952 (±0.016) | 0.693 (±0.094) | 0.493 (±0.076) |
| | | **Naïve Bayes** | 0.160 (±0.026) | 0.558 (±0.082) | 0.898 (±0.023) | 0.575 (±0.081) | 0.537 (±0.069) |
| | | **Neural Net** | 0.152 (±0.025) | 0.658 (±0.077) | 0.885 (±0.025) | 0.602 (±0.075) | 0.586 (±0.063) |
| | | **avNNet** | 0.132 (±0.025) | 0.575 (±0.081) | 0.928 (±0.020) | 0.671 (±0.082) | 0.553 (±0.067) |
| **Trail + Stroop** | **Trail + Stroop** | **Logistic Regression** | 0.189 (±0.033) | 0.500 (±0.094) | 0.895 (±0.029) | 0.638 (±0.099) | 0.515 (±0.082) |
| | | **Linear Discriminant Analysis** | 0.201 (±0.034) | 0.433 (±0.093) | 0.897 (±0.029) | 0.629 (±0.104) | 0.460 (±0.086) |
| | | **SVM (Linear)** | 0.174 (±0.032) | 0.300 (±0.086) | 0.966 (±0.017) | 0.817 (±0.138) | 0.350 (±0.098) |
| | | **SVM (Gauss)** | 0.221 (±0.035) | 0.067 (±0.047) | 0.967 (±0.017) | 0.333 (±0.230) | 0.083 (±0.076) |
| | | **Random Forest** | 0.179 (±0.032) | 0.367 (±0.091) | 0.942 (±0.023) | 0.800 (±0.121) | 0.455 (±0.092) |
| | | **Naïve Bayes** | 0.180 (±0.0032) | 0.533 (±0.094) | 0.895 (±0.029) | 0.610 (±0.097) | 0.534 (±0.081) |
| | | **Neural Net** | 0.201 (±0.034) | 0.600 (±0.092) | 0.853 (±0.034) | 0.511 (±0.087) | 0.512 (±0.076) |
| | | **avNNet** | 0.201 (±0.034) | 0.500 (±0.094) | 0.878 (±0.031) | 0.551 (±0.096) | 0.489 (±0.082) |
| **Trail + Semantic** | **Trail + Semantic** | **Logistic Regression** | 0.152 (±0.027) | 0.433 (±0.066) | 0.951 (±0.018) | 0.742 (±0.104) | 0.503 (±0.082) |
| | | **Linear Discriminant Analysis** | 0.158 (±0.028) | 0.467 (±0.066) | 0.937 (±0.021) | 0.708 (±0.100) | 0.520 (±0.080) |
| | | **SVM (Linear)** | 0.180 (±0.029) | 0.325 (±0.081) | 0.943 (±0.019) | 0.643 (±0.119) | 0.363 (±0.087) |
| | | **SVM (Gauss)** | 0.186 (±0.029) | 0.275 (±0.076) | 0.950 (±0.018) | 0.560 (±0.132) | 0.290 (±0.089) |
| | | **Random Forest** | 0.146 (±0.027) | 0.517 (±0.087) | 0.944 (±0.019) | 0.722 (±0.097) | 0.553 (±0.078) |
| | | **Naïve Bayes** | 0.158 (±0.028) | 0.667 (±0.082) | 0.888 (±0.027) | 0.645 (±0.081) | 0.613 (±0.067) |
| | | **Neural Net** | 0.152 (±0.027) | 0.650 (±0.084) | 0.901 (±0.025) | 0.688 (±0.083) | 0.596 (±0.069) |
| | | **avNNet** | 0.146 (±0.027) | 0.575 (±0.086) | 0.930 (±0.022) | 0.707 (±0.091) | 0.588 (±0.074) |
| **Trail + Walk-12** | **Trail + Walk-12** | **Logistic Regression** | 0.185 (±0.030) | 0.475 (±0.089) | 0.897 (±0.026) | 0.518 (±0.094) | 0.467 (±0.079) |
| | | **Linear Discriminant Analysis** | 0.149 (±0.028) | 0.542 (±0.089) | 0.926 (±0.023) | 0.668 (±0.094) | 0.567 (±0.077) |
| | | **SVM (Linear)** | 0.184 (±0.030) | 0.217 (±0.073) | 0.963 (±0.016) | 0.595 (±0.155) | 0.270 (±0.093) |
| | | **SVM (Gauss)** | 0.191 (±0.030) | 0.117 (±0.058) | 0.978 (±0.013) | 0.500 (±0.270) | 0.147 (±0.089) |
| | | **Random Forest** | 0.143 (±0.027) | 0.450 (±0.089) | 0.955 (±0.018) | 0.722 (±0.103) | 0.516 (±0.084) |
| | | **Naïve Bayes** | 0.299 (±0.035) | 0.717 (±0.080) | 0.696 (±0.040) | 0.378 (±0.061) | 0.490 (±0.063) |
| | | **Neural Net** | 0.210 (±0.031) | 0.567 (±0.088) | 0.845 (±0.031) | 0.478 (±0.061) | 0.500 (±0.073) |
| | | **avNNet** | 0.174 (±0.029) | 0.417 (±0.088) | 0.926 (±0.023) | 0.556 (±0.105) | 0.450 (±0.084) |
| **Demographics** | **All (Trail)** | **LogReg + Majority** | 0.169 (±0.020) | 0.033 (±0.019) | 0.986 (±0.007) | 0.167 (±0.0223) | 0.040 (±0.034) |
| | | **LDA + Majority** | 0.163 (±0.020) | 0.070 (±0.032) | 0.986 (±0.007) | 0.458 (±0.218) | 0.094 (±0.054) |
| | | **SVM(lin) + Majority** | 0.160 (±0.020) | 0.000 (±0.000) | 1.000 (±0.000) | - | 0.000 (±0.000) |
| | | **SVM(Gauss) + Majority** | 0.152 (±0.020) | 0.090 (±0.040) | 0.992 (±0.005) | 0.714 (±0.199) | 0.152 (±0.066) |
| | | **RF + Majority** | 0.149 (±0.0019 | 0.143 (±0.049) | 0.985 (±0.007) | 0.667 (±0.148) | 0.229 (±0.072) |
| | | **Naive Bayes + Majority** | 0.158 (±0.020) | 0.243 (±0.059) | 0.957 (±0.0012 | 0.517 (±0.104) | 0.309 (±0.070) |
| | | **Neural Net + Majority** | 0.172 (±0.021) | 0.143 (±0.049) | 0.958 (±0.012) | 0.338 (±0.115) | 0.172 (±0.065) |
| | | **AvNNet + Majority** | 0.173 (±0.021) | 0.193 (±0.054) | 0.95 (±0.013) | 0.425 (±0.105) | 0.261 (±0.067) |
| **Trail** | **All (Trail)** | **LogReg + Majority** | 0.140 (±0.019) | 0.294 (±0.062) | 0.972 (±0.010) | 0.675 (±0.104) | 0.392 (±0.072) |
| | | **LDA + Majority** | 0.143 (±0.019) | 0.309 (±0.063) | 0.965 (±0.011) | 0.675 (±0.099) | 0.411 (±0.071) |
| | | **SVM(lin) + Majority** | 0.152 (±0.010) | 0.162 (±0.049) | 0.982 (±0.008) | 0.679 (±0.146) | 0.225 (±0.071) |
| | | **SVM(Gauss) + Majority** | 0.146 (±0.019) | 0.223 (±0.056) | 0.979 (±0.009) | 0.620 (±0.123) | 0.296 (±0.073) |
| | | **RF + Majority** | 0.119 (±0.018) | 0.409 (±0.067) | 0.975 (±0.009) | 0.758 (±0.085) | 0.508 (±0.069) |
| | | **Naive Bayes + Majority** | 0.149(±0.019) | 0.412 (±0.067) | 0.939 (±0.014) | 0.638 (±0.083) | 0.472 (±0.066) |
| | | **Neural Net + Majority** | 0.154(±0.020) | 0.449 (±0.068) | 0.926 (±0.016) | 0.597 (±0.077) | 0.483 (±0.063) |
| | | **AvNNet + Majority** | 0.152 (±0.020) | 0.333 (±0.064) | 0.951 (±0.013) | 0.654 (±0.092) | 0.407 (±0.069) |
| **None** | **All** | **Majority** | 0.160 (±0.020) | 0.000 (±0.000) | 1.000 (±0.000) | - | 0.000 (±0.000) |
| **None** | **Trail** | **Majority** | 0.184 (±0.027) | 0.000 (±0.000) | 1.000 (±0.000) | - | 0.000 (±0.000) |